\documentclass[preprint2]{aastex}
\usepackage[varg]{txfonts}
\usepackage{lscape}

\slugcomment{}

\shorttitle{AO images of QSOs}
\shortauthors{Liuzzo et al.}

\begin{document}

\title{MAD Adaptive Optics Imaging of High Luminosity Quasars: \\
A Pilot Project. }

\author{E. Liuzzo\altaffilmark{1}, R. Falomo\altaffilmark{2}, S. Paiano\altaffilmark{2}, A. Treves\altaffilmark{3}, M. Uslenghi\altaffilmark{4}, C. Arcidiacono\altaffilmark{5}, A. Baruffolo\altaffilmark{2}, E. Diolaiti\altaffilmark{5}, J. Farinato \altaffilmark{2}, M. Lombini\altaffilmark{5}, A. Moretti\altaffilmark{2}, R. Ragazzoni\altaffilmark{2}, R. Brast\altaffilmark{6}, R. Donaldson\altaffilmark{7}, J. Kolb\altaffilmark{7}, E. Marchetti\altaffilmark{7} and S. Tordo\altaffilmark{7}}

\affil{$^{1}$ Osservatorio di Radioastronomia, INAF, via Gobetti 101, 40129 Bologna, Italy;\\$^{2}$ Osservatorio Astronomico di Padova, INAF, vicolo dell'Osservatorio 5, 35122 Padova, Italy;\\$^{3}$ Universit\`a dell'Insubria (Como), Italy, associated to INAF and INFN;\\$^{4}$ INAF-IASF, via E. Bassini 15, I-20133 Milano, Italy;\\$^{5}$ Osservatorio Astronomico di Bologna, INAF, Bologna, Via Ranzani 1, 40127 Bologna, Italy;\\$^{6}$ Dipartimento di Fisica e Astronomia, Universit\`a di Bologna, Via Irnerio, 46, 40126, Bologna, Italy;\\$^{7}$ European Southern Observatory, Karl-Schwarschild-Str 2, 85748 Garching bei M\"unchen, Germany.}

\email{liuzzo@ira.inaf.it}
\begin{abstract}
We present near-IR images of five luminous quasars at z$\sim$2 and one at z$\sim$4 obtained with an experimental adaptive optics instrument at the ESO Very Large Telescope. The observations are part of a program 
aimed at demonstrating the capabilities of multi-conjugated adaptive optics imaging combined with the use of 
natural guide stars for high spatial resolution studies on large telescopes. The observations were mostly obtained under poor seeing conditions but in two cases. In spite of these non optimal conditions, the resulting  images of point sources have cores of FWHM $\sim$ 0.2 arcsec. We are able to characterize the host galaxy properties for 2 sources and set stringent upper limits to the galaxy luminosity for the others.
We also report  on the expected capabilities for investigating the host galaxies of distant quasars with adaptive optics systems coupled with future Extremely Large Telescopes. Detailed simulations show that it will be possible to characterize compact (2-3 kpc) quasar host galaxies for QSOs at z = 2 with nucleus K-magnitude spanning from 15 to 20 (corresponding to absolute magnitude -31 to -26) and host galaxies that are 4 mag fainter than their nuclei.
\end{abstract}


\section{Introduction}\label{intro}

Quasars are among the most luminous sources in the Universe and can be therefore detected and investigated at huge 
distances and early cosmic times.
For this reason, their study can be used to probe a variety of phenomena from 
cosmology models to gas content of the intergalactic medium (from intervening absorption systems) and for an understanding of many aspects of the processes of the formation and evolution of galaxies.

The latter issue gained a lot of interest with the discovery that most of massive galaxies host a dormant supermassive black hole (SMBH, Di Matteo et al. 2005; Callegari et al. 2011, and references therein). This means that accretion onto a SMBH, that is the most likely mechanism for the QSO phenomenon, is strictly linked to the whole process of formation of galaxies. The coevolution of the SMBH and their host galaxies therefore represents a fundamental issue in the modern extragalactic studies. While direct dynamical measurements of the mass of the BH can be derived only for the nearest galaxies (see Peterson 2014 for a review), for more distant objects (z$>$0.1, e.g. Koekemoer et al. 2009) it is possible to deduce the BH mass only from the relationships, derived from a small number of nearby galaxies, between the BH mass (M(BH)) and the global properties  of the galaxies (e.g. stellar velocity dispersion, luminosity of the spheroidal component; see e.g. Ferrarese \& Merritt 2000, Gebhardt et al. 2000; Tremaine et al. 2002, Marconi \& Hunt 2003, Haring \& Rix 2004, Gultekin et al. 2009).
In the case of galaxies with active nuclei (as the case of QSOs) it is possible to estimate directly the BH mass from the dynamical properties of the gas that is under the sphere of influence of the central BH. 
This method was applied by various authors to derive M(BH) of several QSOs at various redshift (e.g. McLure \& Dunlop 2004, Shen et al. 2013 and references therein).

The direct measurements of both BH mass and the mass (luminosity) of their host galaxies allow to investigate the relation between these two fundamental ingredients and to probe their evolution with the cosmic time.
While for the M(BH) estimation of the QSO the spectroscopic observation (provided if a sufficient signal-to-noise is reached) is able to gather a measurement of BH mass virtually at any redshift, in the case of the properties of the QSO host galaxies the measurement is increasingly more 
challenging for higher redshift. This is because the contrast between the bright central nucleus and the starlight from the host galaxy critically depends on the size and shape of the latter compared with that of the point spread function (PSF).
Indeed, for a robust decomposition of the nuclear and host galaxy emission of a QSO, it is extremely important to obtain deep images with as narrow as possible PSF in order to reduce the emission contribution of the central source. This can be done either from space based observations, as from the Hubble Space Telescope (HST, see e.g. Disney et al. 1995, Bahcall et al. 1997, McLure et al. 1999; McLeod \& McLeod 2001, Dunlop et al. 2003) or using ground based imaging under excellent seeing conditions (Hutchings et al. 1999, M{\'a}rquez et al. 2001, Lacy et al. 2002, Kotilainen et al. 2009). 
This situation becomes particularly critical for high redshift sources because of the faintness of the starlight emission and/or the very small size of the galaxies. 

Observations from space with HST can indeed provide an excellent narrow PSF but because of the faintness of high z host galaxies they are usually limited by the small collecting area of the telescope and the limited performances in the near-IR 
(e.g. Kukula et al. 2001, Peng et al. 2006).
The ground based near-IR imaging of high-z quasars, obtained with large telescopes under excellent seeing conditions, can compete with space based observations because of the significantly larger collecting power and thus allow to characterize the properties of distant QSOs (e.g. Falomo et al. 2004,  Kotilainen et al. 2009)
To further explore quasar hosts at high-z 
the use of adaptive optics, that produces narrow PSF, is therefore seen as a natural cure for reducing the light of the central source and to emphasize the more extended emission of its host galaxy and their features in the very close environments.

\begin{figure*}[ht!]
\epsscale{2}
\plotone{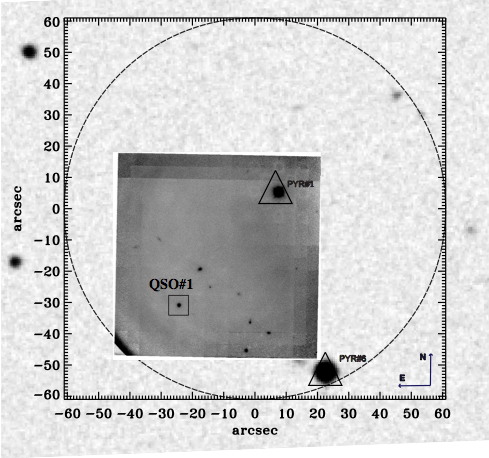} 
\caption{\footnotesize The 2 arcmin field of view for the QSO\#1  (DSS R filter image).
The inner panel is the MAD (CAMCAO) final combined image in Ks filter. The two open triangles show the reference
guide stars used (V=12.3 and 13.4 respectively for Pyr 1 and Pyr 6). The open box indicates the QSO.} \label{fig_star}
\end{figure*}

In the last decade Adaptive Optics (AO, see Davies \& Kasper 2012 for a review) images of QSOs were obtained either to investigate with better spatial resolution the immediate 
environment of low redshift objects (e.g. M{\'a}rquez et al. 2001, 2003; Guyon et al. 2006, Rosario et al. 2011) or to characterize the properties of the host galaxies of high redshift (z $>$ 2--3) quasars (e.g. Croom et al. 2004; Hutchings et al. 1999, 2004; Falomo et al. 2005, 2008; Kuhlbrodt et al. 2005; Wang et al. 2013) . In the latter case the quality of the PSF and its characterization are essential ingredients to disentangle the light of the host galaxy from that of the bright central nucleus.
The large majority of the AO imaging systems requires the presence of a bright star close ($<$ 10-30 arcsec) to the target 
in order to derive the signal of the wavefront perturbation and apply the adaptive corrections. This implies that only the targets that are angularly very close to relatively bright stars could be actually observed with AO systems. 
This limit could be overcome by some present (e.g. Keck, Wizinowich et al. 2006) and future AO instruments using artificial laser guide stars. 
Another approach to cope with this limitation is to use many natural guide stars (NGSs) in the field of view of the target as reference for AO correction and more than one deformable mirror conjugated to different altitudes of the perturbed wavefront. With this Multi Conjugated Adaptive Optics (MCAO) it is possible to obtain a more stable and uniform PSF over the observed field of view and to increase significantly the 
sky coverage even at high galactic latitudes (Moretti et al. 2009, Arcidiacono et al. 2010, Meyer et al. 2011).

In this paper we report on observations of high-z luminous QSOs gathered with Multi-Conjugate Adaptive Optics Demonstrator (MAD) at the  European Southern Observatory (ESO) Very Large Telescope (VLT).
Other examples of nearby active galaxies observed by MAD have been presented by Falomo et al. (2009) and Liuzzo et al. (2011, 2013). 
The selected QSOs are high-luminous sources with V magnitude ranging between 18.6 and 20.1. 
The detection of the host galaxy in those objects appears therefore more difficult with respect to low-luminous sources, because of the presence of very bright central nuclei.
Due to non optimal observing conditions all QSOs but one (observed in the conventional AO mode) were observed in the Ground Layer AO (GLAO) configuration (Marchetti et al. 2007) using NGSs with R magnitude in the range between 9.8 and 15.6. In particular,  in GLAO mode a set of reference stars are available to provide the control of a deformable mirror conjugated to the ground thus allowing the correction of the optical turbulence in the lower part of the atmosphere volume above the telescope.
Although GLAO aims just to a partial correction of the turbulence, it allows for a much larger compensated FoV with respect to the conventional single conjugate AO mode (SCAO) which uses only one reference star in the field. In spite of the modest observing sky conditions, we are able to resolve the host galaxy of two objects and to set upper limits to the host luminosity for the others.
Finally, future perspectives for the capabilities in this field with European Extremely Large
Telescope (E-ELT) imaging are presented.

Throughout this work we use H$_{0}$ = 70 km s$^{-1}$ Mpc$^{-1}$,  $\Omega_{\rm m} =
0.3$, and  $\Omega_\Lambda = 0.7$.

\begin{table*}[ht!] 
\caption{Journal of the MAD observations.}\label{tab_MADobs}
\footnotesize
\centering
\begin{tabular}{|llcccccccccl|}
\hline
ID 	& Object    &      RA          &   DEC      &   z     & V       &   AO  & seeing    & expt  & FWHM & N$_{*}$ &  Dist \\
        & name      &     (J2000)      &  (J2000)   &         & mag     &mode            &   arcsec    & min   & arcsec  &                         & arcsec\\ 

\hline\hline
QSO\#1 & ICS96 001721.4-391641        &     00 19 50.7    & -39 00 02 &  2.120   & 19.33   &  GLAO     &    1.24     & 75    & 0.17 &1 & 49 \\
QSO\#2  & 2dFGRS S505Z169             &     01 13 39.7   & -33 43 26 &  1.550   & 18.76   &  SCAO      &   1.67     & 51    & 0.30 & 1 & 21   \\
QSO\#3 & PKS 0227-369                 &      02 29 28.4   & -36 43 57 &  2.115  & 19.00    &   GLAO     &    1.45     & 80    & 0.23 & 2 & 13, 50 \\
QSO\#4 & 2QZ J025907.2-313412         &     02 59 07.2    & -31 34 12 &  2.235  & 19.75   &   GLAO     &    1.90     & 30    & 0.25  & 1 & 21\\
QSO\#5 &  BR J2017-4019               &    20 17 17.1     & -40 19 24 &  4.131  & 18.57   &  GLAO      &    0.57     & 40    & 0.15 & 2 & 24, 26\\
QSO\#6 & 2QZ J223006.0-281017         &      22 30 06.0   & -28 10 17 &  2.401  & 20.11   &  GLAO     &     0.83     & 35    & 0.19 & 2 & 14, 21\\
\hline 
\end{tabular}
\tablenotetext{}{
\footnotesize Col.s 1--2: ID and object name of our targets; col.s 3--4 RA and DEC in J2000; col. 5: redshift; col.6: source apparent V magnitude from NED; col.7: observing AO mode; col. 8: seeing during the observations as derived by the ESO DIMM in the V band (rescaled at the Zenith); col. 9: total integration time; col. 10: FWHM of the PSF core; col. 11: number of stars used for the PSF fit; col. 12: : angular distance of the star(s) from the QSO.}
\end{table*}

\section{MAD Observations} \label{sect_MADobs}

We performed Ks-band observations of 6 high-z luminous QSOs using the ESO MAD, mounted at UT3 (Melipal) of VLT.
MAD was an experiment devoted to demonstrate the feasibility of the Multi Conjugate Adaptive Optics (MCAO) reconstruction technique as a test-bench for the E-ELT and the 2nd Generation VLT Instruments. MAD was designed to perform adaptive optics correction in J, H and Ks bands over 2 arcmin (Fig. \ref{fig_star}) on the sky by using relatively bright (m$_{v}<$14) NGS. 
We refer to Marchetti et al. 2003 for the detailed description. In this paper, we consider MCAO correction Layer Oriented Multi-Pyramid Wave Front Sensor observations (Ragazzoni 1996, Ragazzoni et al. 2000, Marchetti et al. 2005). The detector (Amorim et al. 2004, 2006) has a 57$\arcsec\times57\arcsec$ field of view (FoV) with pixel size of 0.028 \arcsec.

Images of the targets were obtained following a 5-position jitter pattern, with offsets of 5 arcsec, to ensure an adequate subtraction of the computed sky background.
For each observation, we constructed a reference sky image, which was subtracted from the science frame to obtain the net intensity frame.  
Finally, we combined all sky-subtracted and registered frames into a final image. 

The list of the observed QSOs and the journal of the observations are presented in Table \ref{tab_MADobs}. The seeing measured by the ESO DIMM in the V band (rescaled at the Zenith) during the observations
varied from 0.6\arcsec to 1.9\arcsec, with an average of 1.2\arcsec. 
Only in two cases, it was better than 1\arcsec and in spite of the non optimal seeing, the high performance of MAD allows to obtain images of point like sources with cores of full width half maximum (FWHM) $\sim$0.2\arcsec.

In addition to the image of the quasars, we detected in the fields a number of faint and very small galaxies. We discuss them in the Appendix \ref{App}. Despite the modest seeing conditions, we are able to characterize their morphologies confirming the final good performance of these MAD observations. 

\section{Image analysis}

In order to derive the properties of the galaxies hosting the QSOs, we performed a two-dimensional fit of the MAD images of the QSOs using the Astronomical Image Decomposition and Analysis (AIDA) package (Uslenghi \& Falomo (2008)), an IDL-based software used also in our previous studies of QSO host galaxies (Falomo et al. 2008, 2014; Kotilainen et al. 2007, 2009; Decarli et al. 2012). AIDA is specifically designed to provide a simultaneous decomposition into the nucleus and the surrounding nebulosity: the nuclear component is described by the local PSF of the image, while for the host galaxy we assumed a galaxy model described by a Sersic profile convolved with the proper PSF. 

\begin{figure*}[ht!]
\begin{center}
\epsscale{2.2}
\plotone{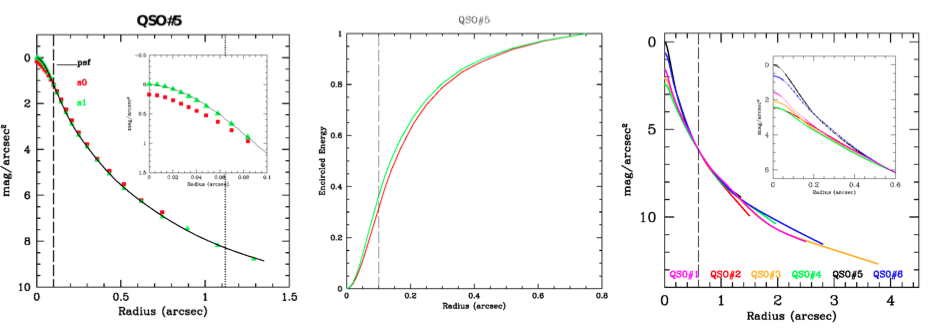}
\caption{\footnotesize {\it Left:} Comparison among the radial brightness profiles of the two stars (green triangles and red squares) used for the PSF fit (colors) and the PSF fit (black solid line) for QSO\#5. The dashed and dotted line represent respectively the inner radius and maximum radius used for the PSF fit.  {\it Center}: Encircled Energy distribution of the two QSO\#5 stars.  {\it Right}: Comparison of the radial brightness profile of the PSF for the 6 observed fields (solid color lines). The PSF are very similar at radii larger than 0.6 arcsec while they differ significantly in the inner core (see inset) due to different observing conditions and AO correction (see text). The dashed line indicates the radius at which all PSFs are normalized.} \label{figPSFall}
\end{center}
\end{figure*}

\subsection{Point spread function} \label{sect_psf}

\begin{figure*}[ht!]
\begin{center}
\epsscale{2.2}
\plotone{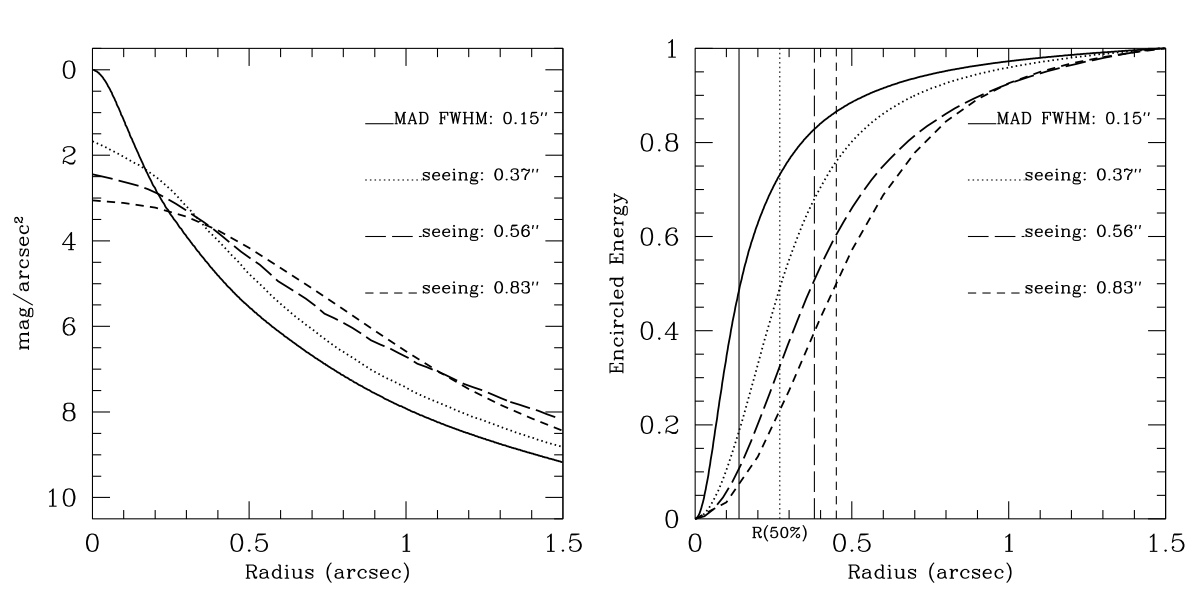}
\caption{\footnotesize Comparison among PSF average radial brightness profile ({\it left panel)} and the encircled energy distribution ({\it right panel}) of AO (MAD for QSO\#5) and non AO images obtained under different seeing conditions (VLT--ISAAC) with dotted and long dashed lines, and ESO 2.2 with short dashed line. All PSFs are normalized to the same total flux. In the {\it right panel}, the half light radius R(50$\%$) for each PSF is indicated with vertical lines. The effects of the seeing in the PSF shape in the VLT--ISAAC observations are evident: high values of seeing produce more flattening in the PSF core and wing and the half of the PSF total energy is contained within higher radius (R(50$\%$))} \label{fig_VLTMAD}
\end{center}
\end{figure*}

To disentangle the extended host galaxy light from the nuclear source, the key factors are the nucleus-to-host magnitude ratio and the seeing (the latter affecting the shape of the PSF). The most critical aspects of the image decomposition are thus the determination of a PSF and the estimate of the background level around the target, which may strongly affect the faint signal from the host galaxies. To model the PSF shape in each frame, the first step is to select the suitable field stars based on their FWHM, sharpness, roundness, and signal--to--noise ratio. 
The field of our QSOs are generally poor of stars. We were able to use one or two stars for field at typical distance between 15--50 arcsec from the target (see Table \ref{tab_MADobs}).
For each star, the region used for PSF fit was selected by defining an internal and an external radius of a circular area around each star. Setting the internal radius to a non--zero value allows the exclusion of the core of bright, saturated stars. We ensured that the external radius of the star used for the PSF extends more than that of the QSO. The local sky background is computed in a ring centered on the source, and its uncertainty was estimated from the standard deviation of the values computed in sectors of concentric sub-annuli included in this area.
All extra sources, saturated pixels and other defects affecting the images found inside these areas were masked out with an automated procedure. The PSF model was obtained using a multi function 2D model composed by 3 Gaussians (representing the inner region of the PSF) and/or one exponential function (representing the extended wing).

We determine the fit of the PSF for each field. 
In each frame, the PSF is expected to change in the central region (radii $<$0.1 arcsec) as a function of the position in the frame of the star(s) used to model the PSF. 
As example we show, in Fig. \ref{figPSFall}, the radial brightness profiles of the two stars used for the PSF fit for QSO\#5 compared to the PSF model.  In the inner 0.1 arcsec region of the PSF,  the radial profiles of the 2 stars is slightly different and the fit of the PSF is dominated by the brightest star. The difference in terms of encircled energy for these two stars is $\sim$ 5\% at radii $<$ 0.1 arcsec (see Fig. \ref{figPSFall}).
A better discussion of the PSF variations in the field is given in the Appendix \ref{App}.

The comparison of the PSF models obtained for different targets and  in different nights is also shown in Fig. \ref{figPSFall}. As expected, they differ in the central region, at radii $<$ 0.6 arcsec. This is due to different seeings of each observation, adaptive optic corrections applied and the position in the frame of the stars used for the PSF model.  In particular, we note that the radial profile of the PSF fit for the QSO\#5 field, which has the best seeing (0.57 arcsec, Table \ref{tab_MADobs}), is the most narrow and peaked in the center, while the PSF fit for the QSO\#4 field, which has the highest seeing (1.9 arcsec, Table \ref{tab_MADobs}), has the broadest inner radial profile. Finally, at large radii $>$ 0.6 arcsec, the PSF fits have very similar radial profiles. This further ensures that the shape of the PSF at radii between 0.6 and 2 arcsec is stable and marginally dependent on the observing conditions.

In Fig. \ref{fig_VLTMAD}, we compare the radial brightness profile of the best PSF fit (and its encircled energy distribution) obtained in this work for the QSO\#5 field with the PSF fits of non-AO observations obtained with different values of seeing. 
The high quality of the MAD PSF is clear: even with a non-optimal seeing ($\sim$0.6 arcsec), the MAD AO corrections produce a narrower and better peaked PSF than the best non-AO observations and the R(50$\%$) is the smallest one. To obtain a narrow as possible PSF is one of the key factor to disentangle the extended host galaxy emission from the nucleus. 

\subsection{The host galaxy characterization} \label{sect_host}

In order to discriminate between resolved and unresolved targets, we first fit the images of our sources with the pure PSF model. As the PSF fit is not well defined in the core because of positional dependence over the field (see Sect. \ref{sect_psf}) we excluded from the QSO fit the region at radii $<$0.1 arcsec. In the cases where an excess is observed in the residuals, we performed again the fit model with a point-source (the nucleus) plus a Sersic law (describing the host galaxy) convolved to the PSF model. 
The fit of the QSO image was performed using a $\chi^2$ minimization assuming a noise model that includes the statistical noise from the source and the uncertainty in the background subtraction. The latter was obtained from the comparison of the level of the background in various annuli around the targets at radii around 2 and 4 arcsec from the target. The uncertainty of the PSF at large radii is negligible with respect to these errors since the stars are much brighter.

For the unresolved objects we evaluated the upper limit of the host galaxy luminosity (Tab. \ref{tab_host}) by adding the flux of a galaxy assuming a De Vaucouleurs model with an effective radius R$_{eff}$ of 5 kpc to the object until the $\chi^{2}$ of fit to these data become 20\% worse than that obtained from the fit with the scaled PSF alone. The input half light radius in arcsec was derived assuming the redshift of the QSO.


\begin{figure*}[ht!]
\epsscale{1.035}
\plotone{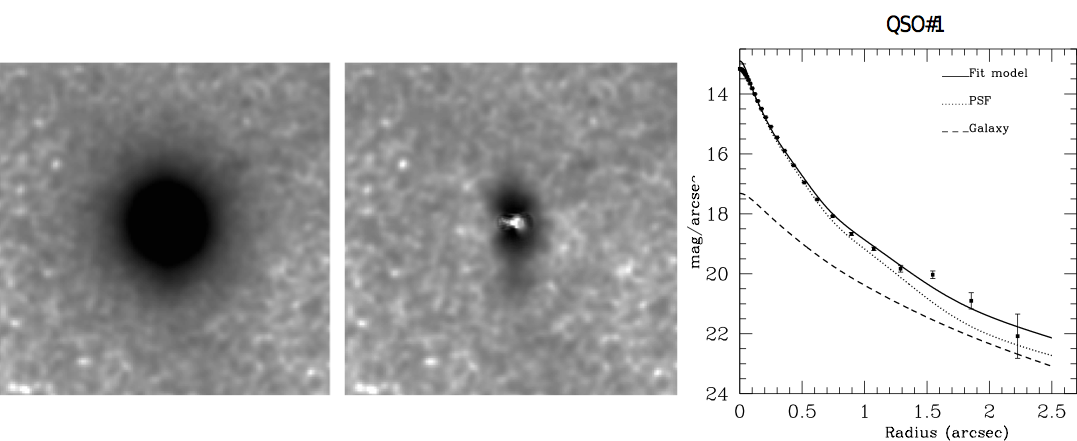}
\epsscale{1.071}
\plotone{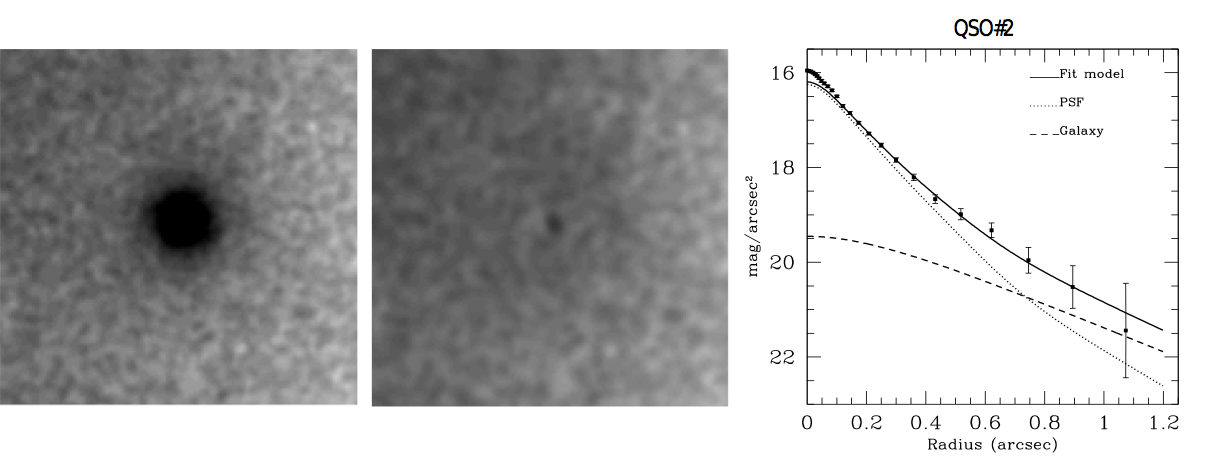}
\plotone{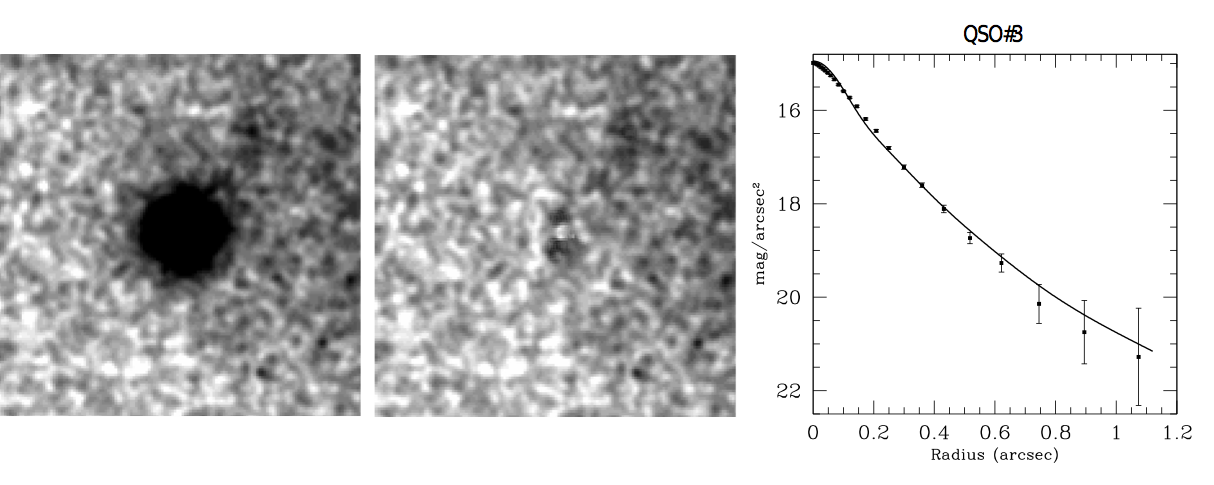}
\plotone{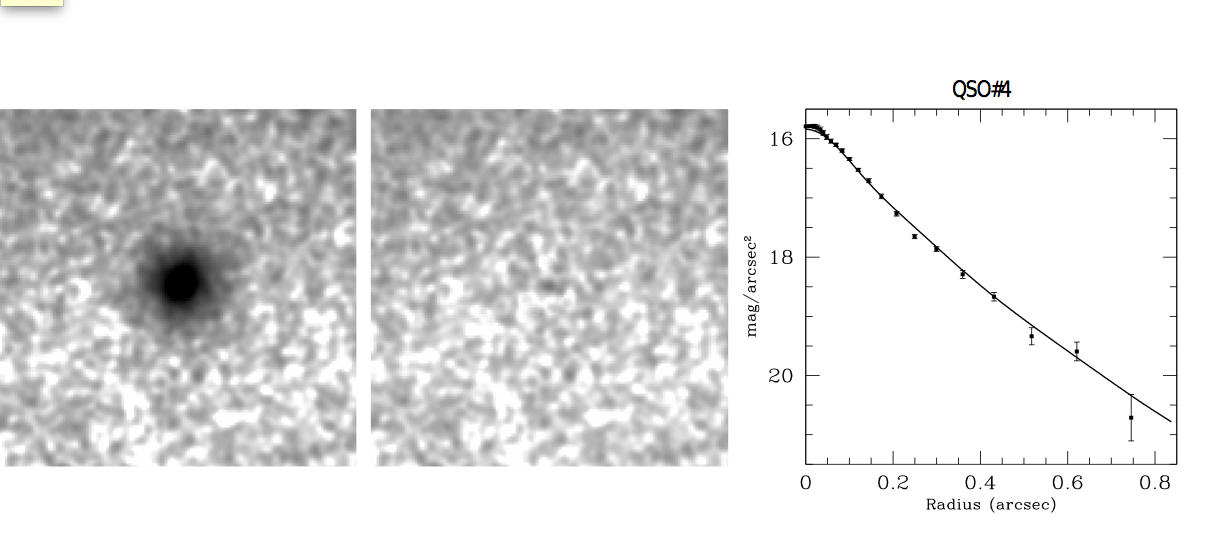}
\plotone{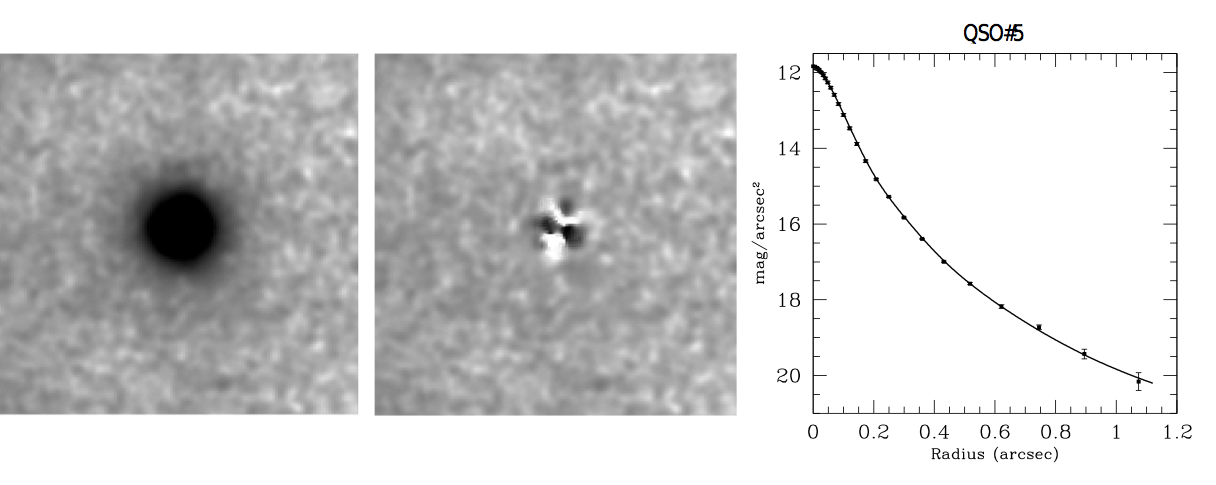}
\plotone{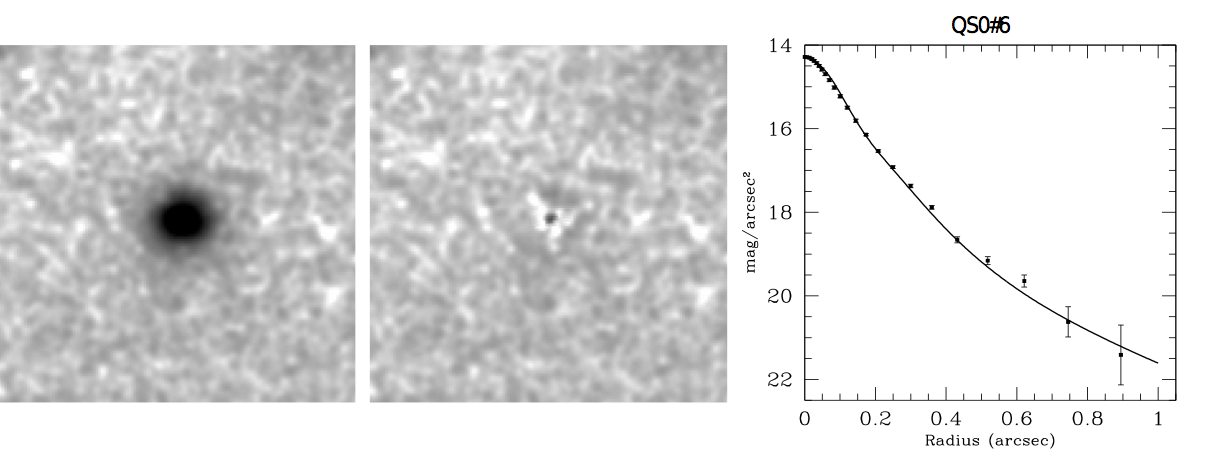}
\caption{\footnotesize {\it Left panels}: Ks images obtained for the QSOs of this work. {\it Middle panels}: PSF subtracted images. {\it Right panels}: radial surface brightness ({\it squares}), with overplotted the best-fit model ({\it solid line}) which consists for the resolved objects by a point source ({\it dotted line}) plus a Sersic profile host galaxy with n=5 for QSO\#1 and n=1 for QSO\#2 convolved with the PSF ({\it dashed line}). The errors bars associated to the radial profiles take into account the statistical errors of the signal of the QSO and of the background. }\label{fig_ImageAll} 
\end{figure*}

\begin{table}[ht!]
\caption{QSOs host galaxies properties} \label{tab_host}
\scriptsize
\centering
\begin{tabular}{|lllrlr|}
\hline
Object&   z               &  m(K)     & m(K) & M(R)     & M(R)\\
 name &                   &  nucleus  & host & nucleus  & host\\
\hline\hline
QSO\#1   &  2.12          &15.2     & 17.4         &    -26.9         &-25.61   \\
QSO\#2   &  1.55          & 17.9   &  18.8        &    -26.5          &-25.27   \\
QSO\#3  &  2.11           & 16.8    & $\geq$18.5   &    -26.3        & $\geq$-24.55 \\
QSO\#4  &  2.23           & 17.5   & $\geq$19.2    &    -25.7         & $\geq$-23.93   \\
QSO\#5  & 4.13            & 15.2   & $\geq$17.7    &    -29.8         & $\geq$-27.30   \\
QSO\#6  &  2.40            &17.0     & $\geq$18.8   &    -26.3         & $\geq$-24.50 \\
\hline
\end{tabular}
\tablenotetext{}{
\footnotesize Col.1: object name;  col. 2: redshift; col.s 3-4: total and host galaxy}
\tablenotetext{}{
\footnotesize  Ks-band apparent magnitude; cols. 5--6: nucleus and host galaxy rest }
\tablenotetext{}{
\footnotesize frame R absolute magnitude. For unresolved sources, upper limits of}
\tablenotetext{}{
\footnotesize the host galaxy luminosity are given.}
\end{table}

\section{Results} \label{sect_results}

The MAD images of the 6 QSOs are reproduced in Fig. \ref{fig_ImageAll} together with the PSF subtracted images and radial brightness profiles plus the best fit model.
 In 2 out of 6 cases (QSO\#1 and QSO\#2), the best-fit model comprises a point source plus an extended component, corresponding to a Sersic profile host galaxy with a Sersic index n=5 and n=1 for QSO\#1 and QSO\#2 respectively. In 4 out of 6 QSOs the model fit composed by a pure PSF well represents the radial profile of the object. Therefore these targets are unresolved.

\begin{figure}[t!]
\epsscale{1.08}
\plotone{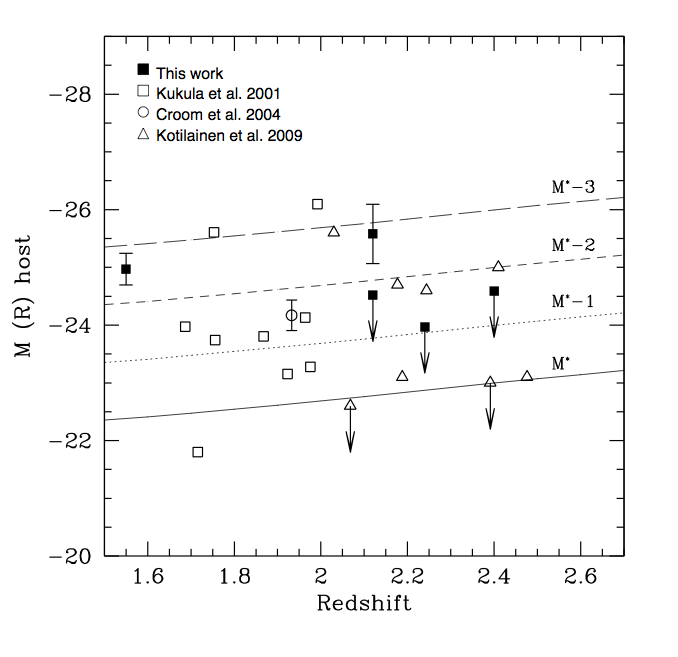} 
\caption{\footnotesize Host galaxy absolute magnitudes in R band for the QSOs of this work (filled squares) compared with available literature data of Kukula et al. 2001 (open squares), Croom et al. 2004 (open circle), and Kotilainen et al. 2009 (open triangles) in the 1.5$<$z$<$2.5 redshift range. The arrows represent the upper limits to the host galaxy luminosities. The solid, dotted and dashed lines show the values of M$^{*}$,M$^{*}$-1, M$^{*}$-2, and M$^{*}$-3  where M$^{*}$ is the $R$-band characteristic absolute magnitude of the Schechter luminosity function undergoing passive evolution (M$^{*}_{z=0}$=-21.17).}\label{fig_zMRH}
\end{figure}

To compare the properties of the quasar hosts with previous detections at similar redshift, we transformed the apparent Ks band magnitude into absolute magnitude in R band (Tab. \ref{tab_host}), as the Ks band at redshift 1.5--2.5 is close to R band rest frame. To perform the color and $k$--correction transformations, we assumed an elliptical galaxy template (Mannucci et al. 2001) for the host galaxy, and a composite quasar spectrum (Francis et al. 1991) for the nucleus. 
In Fig. \ref{fig_zMRH}  we compare the host galaxy magnitudes and upper limits derived in this paper with those available from previous works of Kukula et al. 2001, Croom et al. 2004, and Kotilainen et al. 2009 in the same redshift range. All literature observed magnitudes are converted to the absolute R band, similarly to our sample, and adopting the same cosmological framework.  

The two resolved QSOs have host galaxies with absolute magnitude M(R)$\sim$ -25.5, while upper limits are lower than -24.5. In all cases, they correspond to a range between M$^{*}$ and M$^{*}$ -3 where M$^{*}$ is the $R$--band characteristic luminosity of the Schechter galaxy luminosity function in the redshift range 1.5--2.5, evolved from the local M$^{*}_{z=0}$ = -21.17 (Nakamura et al. 2003).

\section{Notes to individual objects} \label{sect_notes}

{\bf QSO\#1 -- [ICS96] 001721.4-391641} This radio quiet QSO is at $z=$2.12. 
It is marginally resolved.  The best fit parameters for the host galaxy are K(host) = 17.4, R$_{eff}=$8.3 kpc, and the core surface brightness K(nucleus)=15.2. 

{\bf QSO\#2 -- 2dFGRS S505Z169:} The literature redshift (NED) of this object is 3.00. We reanalyzed the 2dF Galaxy Redshift Survey (2dFGRS, Colless et al. 2003) spectrum finding emission lines at 3960, 1540, 7150$\AA$ that we interpreted as emission lines of CIV, CIII, MgII of an object at z$\sim$1.55. This is the nearest radio quiet QSO of the sample being at $z$=1.55.  This is the only observed case in SCAO mode. 
The best fit parameters for the host galaxy are K(host) = 18.8, R$_{eff}=$3.4 kpc, and the core surface brightness K(nucleus)=17.9.

{\bf QSO\#3 -- PKS 0227-369:} It is a $z$= 2.115 source and it is among our small sample the only radio loud QSO (Garofalo et al. 2013). Its morphology is point-like from the arcsec to parsec scales (e.g. Condon et al. 1998, Murphy et al. 2010, Kovalev et al. 2009) with a total flux density of $\sim$ 120 mJy at 1.4 GHz in NVSS image. It is also a gamma-ray source detected with Fermi (source 3FGL J0229.3-3643, e.g. Ackermann et al. 2011, 2013; Lyu et al. 2014). 
This target is non resolved with K(nucleus) = 16.8.

{\bf QSO\#5 -- BR J2017-4019:}  It is the QSO with the highest redshift ($z$=4.131) of our targets. It is a radio quite object exhibiting strong intrinsic absorption. The  Ly($\alpha$),  C IV and Si IV emission lines are completely absorbed (Storrie-Lombardi et al. 2001). 
From our data, this QSO is unresolved with K(nucleus) = 15.2. 

\section{Perspectives on future ELTs} \label{sect_ELT}

In the next decade the future ground based extremely large (30-40m aperture) telescopes are expected to become operative (e.g. E-ELT\footnote{http://www.eso.org/sci/facilities/eelt/}, TMT\footnote{http://www.tmt.org/}, GMT\footnote{http://www.gmto.org/}). This new generation of telescopes, combined with AO system (mainly assisted by laser guide stars), will achieve unprecedented combination of high spatial resolution and sensitivity. 
In this section we briefly exploit the expected capabilities of E-ELT for the study of QSO host galaxies. To this aim we adopt the baseline parameters of the phase A study of one of the first light instrument MICADO (Multi-AO Imaging Camera for Deep Observations, Davies et al. 2010) that is optimized for imaging close to the diffraction limit of the telescope with a PSF core FWHM in the J (K) bands of 6 (11) mas, respectively. 

Using the Advanced Exposure Time Calculator (AETC v.3.0\footnote{http://aetc.oapd.inaf.it}, Falomo et al. 2011), we produced simulated images of high-z QSOs with MICADO@E-ELT in Ks band. The simulations were performed assuming a primary mirror with a diameter of 39 meters, 28\% obscuration, a read-out noise of 5\textit{e}$^{-}$ and a plate scale of 3 mas. The near-IR sky background for Cerro Paranal is assumed, including the contribution for the thermal emission.  We assumed the last version PSF of the multi conjugate adaptive optics post focal relay as provided by MAORY official website\footnote{For details, see http://www.bo.astro.it/maory/Maory/DATA.thml} and calculated for a 0.6\arcsec seeing.

\begin{figure}[h!]
\epsscale{1}
\plotone{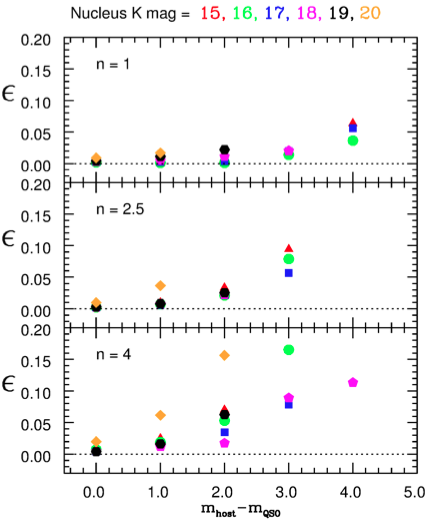} 
\caption{\footnotesize The error budget ($\epsilon$) evaluated for the QSO host galaxy magnitude as function of the difference between the nucleus and host magnitude (see text for details). Each point has been derived from 20 images. 
Each shape-marker and color is referred to a given magnitude of the nucleus which can assume the following values: 15, 16, 17, 18, 19 and 20, with the corresponding exposure time (in minutes) of 20, 20, 30, 40, 80 and 80 respectively.
}\label{fig_sim1}
\end{figure}

In order to explore the QSO observations with different nucleus to host brightness ratio combinations, we varied the apparent magnitude of the nucleus from 15 to 20 in the Ks-band, and for host galaxy we used values from the magnitude of the nucleus to 23. 
Assuming a redshift z$\sim$2, these values correspond to QSOs with nuclei of absolute magnitude between -31 and -26 in the Ks-band and with magnitude of -23 for the fainter host galaxies. Moreover for z$\sim$2 the quasar host galaxies are angularly very small with effective radius of $\sim$0.3\arcsec.  

Each simulated image was created as stack of several images with an exposure time of 1 sec, for a final total exposure time of 20 minutes for QSO nucleus magnitude of 15 and 16, of 30 and 40 minutes for the cases with magnitude 17 and 18 respectively, and finally 80 minutes for QSOs with magnitude of 19 and 20. 
Under these conditions the S/N ratio of the simulated QSO image is roughly constant.

The simulated QSO images were analyzed with the two-dimensional fitting package GALFIT (Peng et al. 2002), an image decomposition program designed to extract the structural components from images of galaxies and/or more complex objects. 
We considered the QSOs as composed by two components: a bright nucleus, fitted by a pure stellar PSF function, plus a host galaxy parameterized by a two-dimensional Sersic profile. For each simulated image, we measure the main photometrical and structural parameters: the total nucleus and host magnitude, the effective radius R$_{eff}$, and the Sersic index \textit{n}. For the PSF we adopted a simulated image of a bright star.
For each nucleus-host combination we performed 20 simulations in order to account for the statistical noise and then derive the median values of the fitting parameters and their 1-sigma dispersion value. 
In Fig.\ref{fig_sim1} we report the results obtained for the estimation of the host galaxy magnitude as function of the difference of magnitude between the nucleus and the host galaxy, for three different values of Sersic index. 
We plot the error budget  ($\epsilon$)  defined as the quadratic combination of the 1-$\sigma$ dispersion value and the difference between the median value of the best-fit host galaxy magnitude parameter and the input value of the same parameter. Therefore this error budget takes into account both the statistical and the systematic errors.

As expected, $\epsilon$ increases for higher nucleus magnitude and with the increase of the difference between the host and nucleus luminosity. Under these conditions it will be possible to determine the host galaxy luminosity of distant (z$\sim$2) QSOs with an accuracy better than 10-15\% for QSOs for which their host galaxy is 3-4 magnitudes fainter than the nucleus.
Moreover, following the same procedure adopted for the magnitude described above, we found that R$_{eff}$ and the Sersic index $n$ of the host galaxies can be determined with an accuracy better than 10\%.  

\begin{figure*}[h!]
\epsscale{2}
\plotone{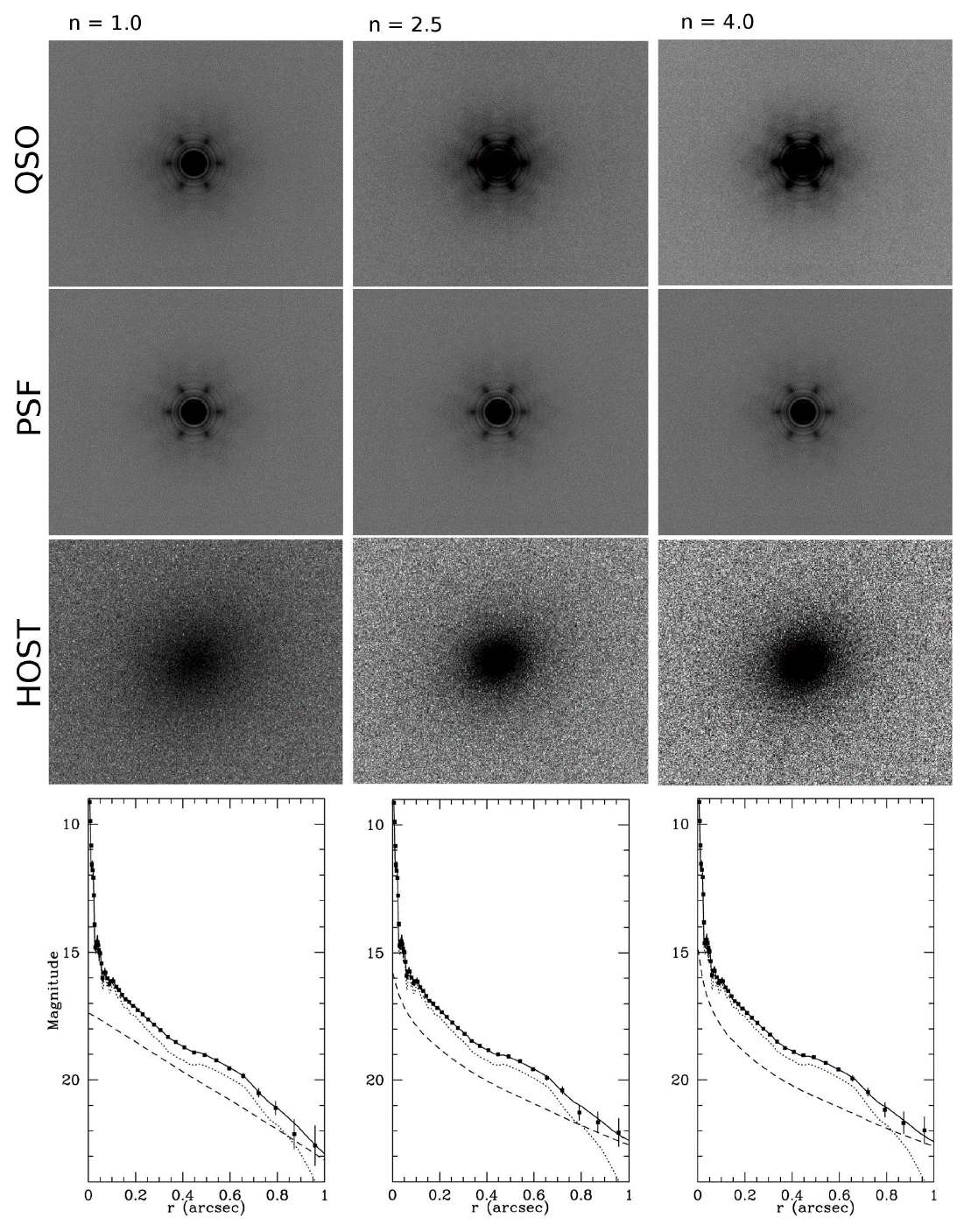} 
\caption{\footnotesize Example of simulated K-band images of QSO with $m_{K}$=17 and its host galaxy with $m_{K}$=19, $R_{eff}$=0.3 \arcsec and Sersic index $n$= 1 (\textit{left column}), $n$=2.5 (\textit{middle column}) and $n$=4.0 (\textit{right column}). From left to right, the panels show the simulated QSO + host galaxy images (\textit{first row}), the PSF corresponding to the nucleus (\textit{second row}) and the PSF subtracted residual image (host galaxy, \textit{third row}). 
The \textit{bottom panels} show the average radial brightness profile of the QSO (filled black squared) compared with the QSO fit (solid line) of the two components model: point nucleus (dotted line) and host galaxy (dashed line) as derived from GALFIT.} \label{fig_sim2}
\end{figure*}

Hence this kind of observations, performed by the high sensitivity and spatial resolution E-ELT camera in combination with the MCAO correction, will allow to well derive with high accuracy the properties of the high \textit{z} QSO host galaxies that are compact objects (2-3 kpc) and are characterized by a value of magnitude until 4 times fainter than their nuclei.  In Fig. \ref{fig_sim2}, we show an example of three QSO simulated images with different host morphologies (Sersic index $n$=1, 2.5, and 4), and the host galaxy that is two magnitudes fainter than their nuclei.

\section{Summary and conclusions}
We presented AO imaging observations in Ks band of five high--luminosity quasars at z$\sim$2 and one ar z$\sim$ 4 obtained with MAD, an experimental AO system at ESO-VLT. Despite the modest seeing conditions for most of these observations but for two cases, the improvement in the PSF obtained by the MAD AO system with respect to the non-AO observations was significant (Sect. \ref{sect_psf}), with a typical FWHM of the PSF core of $\sim$ 0.2 arcsec. In 2 out of 6 cases (QSO\#1 and QSO\#2), we are able to resolve the host galaxy.
In the remaining 4 out of 6 objects, the QSOs are unresolved and we derived upper limits for the host galaxy M(R) $\geq$ -24.5. 

These results demonstrate that deep ground based near-IR images, secured with AO systems, of distant QSOs are able to properly investigate the host galaxies of active nuclei and probe their evolution. Because of the faintness of high--z sources a major impact in this field is expected by the use of the next generation of extremely large telescopes as E-ELT planned to be in operation in the next decade.

\appendix
\section{Appendix} \label{App}
\subsection{Analysis of the PSF variations in the field. } 

The usable field of view of the MAD images is relatively small ($<$ 1 arcmin) thus for high galactic latitude observations there are only few stars in the field that can be employed to evaluate the variation of the PSF with the position in the frame (see Sect. \ref{sect_psf}).
In order to better evaluate  the typical PSF  variations in the field of view of MAD we used a K band image of the globular cluster NGC 6388
(see Moretti et al 2009 for details ) that was secured during the same observing run as the data on QSO presented in this work.

\begin{figure*}[ht!]
\centering
\epsscale{0.7}
\plotone{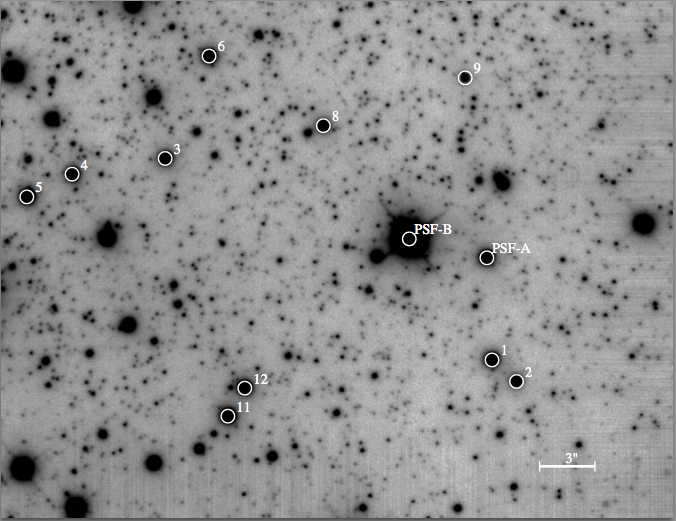}
\caption{K band image of globular cluster  NGC 6388 field obtained by MAD@VLT (see Moretti et al. 2009 for details). The stars used for the analysis of the PSF spatial variations are indicated with circles. Stars with label PSF-A,B were used to compute the PSF model. }
\label{field6388}
\end{figure*}

\begin{figure*}[ht!]
\centering
\epsscale{1}
\plotone{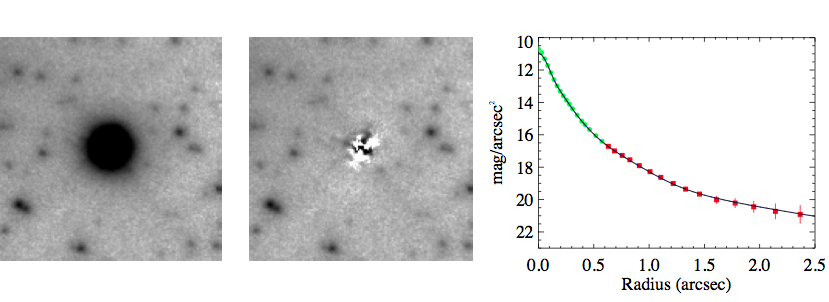}
\caption{Image of star PSF-A  (see Fig. \ref{field6388} ) used for the PSF modeling ({\it left panel}) , the residual image of star  PSF model 
({\it middle panel}) and the average radial brightness profile  ({\it right panel}) of the star (filled green circles) compared with the PSF model (black solid line). The red points are derived by the average radial brightness profile of the brighter star PSF-B  for a suitable characterization of the PSF wings (see text). }
\label{psfstar}
\end{figure*}

\begin{figure*}[ht!]
\epsscale{0.9}
\plotone{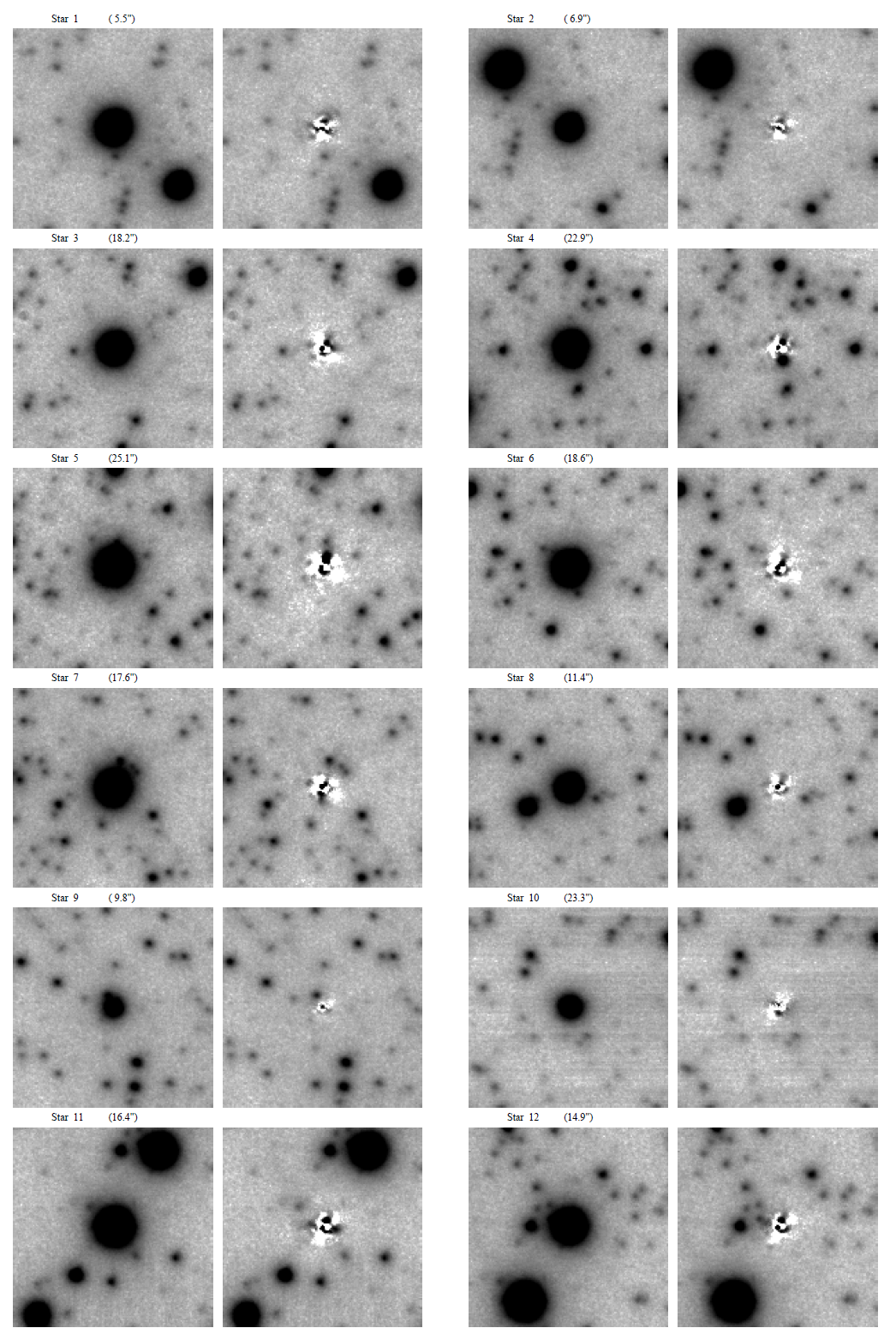}
\caption{The two panels show the residuals ({\it right columns}) of the subtraction of the fitted PSF model (see Fig. \ref{psfstar}) for a dozen  of stars  ({\it left columns})  in the field (see Fig. \ref{field6388}). For each star the distance from the  star PSF-A  is given in parenthesis. 
}
\label{starsubpsf}
\end{figure*}

In this field (see Fig. \ref{field6388}) there are several stars in the range of magnitude between K $\sim$ 13 and K $\sim$ 21. 
We selected 12 relatively bright stars to investigate the spatial variations of the PSF. We excluded few bright stars to avoid saturation and also faint stars because of low signal to noise ratio. Moreover because of the high surface density of objects we selected the stars that are less contaminated by close (fainter) companions. 
We then used two stars  star (PSF-A, B in Fig. \ref{field6388})  in the field to compute the PSF  using
a model similar to that adopted in the analysis of the QSOs. 
The fainter star (PSF-A) is used to fit the core of the PSF while brighter star PSF-B accounts for the  extended wings. The comparison between the stars and the PSF model is shown in Fig. \ref{psfstar}. 
For each selected star we fitted the PSF model after masking out all sources in the field except the selected star (see Fig. \ref{starsubpsf}). The  residual images of the selected stars (see Fig. \ref{starsubpsf}) is finally obtained from the subtraction of the fitted PSF model  from original image of the stars. In order to evaluate quantitatively the residuals, we computed the average residual in three annular regions:  
r $<$ 0.1$^{\prime\prime}$, 0.1$^{\prime\prime}$$<$  r $<$ 0.5$^{\prime\prime}$, and $>$ 0.5$^{\prime\prime}$ r $<$ 1.0$^{\prime\prime}$.  

We found that the average residuals in the three annuli are 6\%, 4\% and 10\%, respectively. Another way to evaluate the variations of PSF is to 
compute the residuals of the azimuthally averaged of the radial brightness profiles of the difference between the selected star and the fitted PSF. In this way we can extend the comparison at larger radii because of the significant  improvement of  the signal-to-noise due to the azimuthal average.
In conclusion, it turns out that for this MAD image the PSF is well stable over a field of view of 30$^{\prime\prime}$.

\clearpage

\begin{figure}[ht!]
\epsscale{1}
\plotone{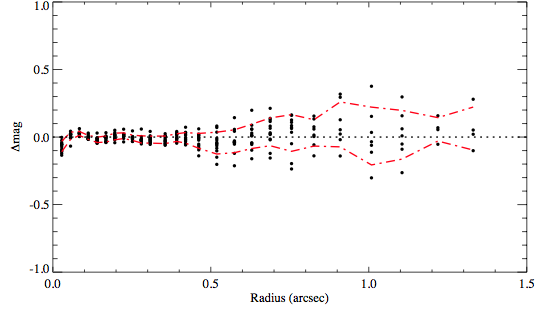}
\caption{Azimuthal average of the radial brightness profiles residuals (filled circles). The dashed red lines represent the rms of the  difference between the radial brightness profile of the selected stars and the fitted PSF model. 
}
\label{residsubpsf}
\end{figure}

\cleardoublepage
\subsection{Faint galaxies in the fields} 

In addition to the image of the quasars, we detected in the QSOs fields a number of faint and very small galaxies.  In order to further exploit the capabilities of these AO images, we briefly report here on the characterization of some of these galaxies through 2D analysis.

\begin{figure*}[ht!]
\epsscale{1}
\plotone{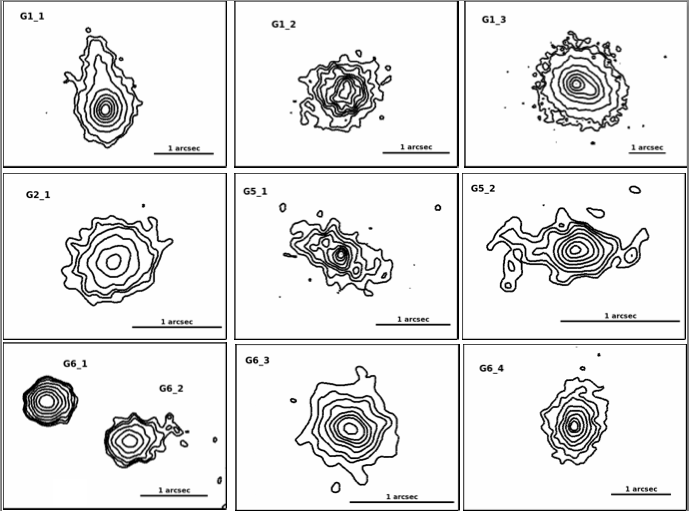} 
\caption{Contour plots of selected galaxies in the fields of QSOs. Sources are named G{\it n}\_{\it a} where {\it n} represents the number of the QSO field and {\it a} is the id number of the object in each field. Some galaxies clearly show a complex morphologies (e.g. G1\_1 and G1\_3). For the other isolated objects, fundamental parameters are given in Tab. \ref{tab_galaxies}. }\label{fig_gal}
\end{figure*}

In Fig. \ref{fig_gal}, we show the contour plots of selected galaxies: some of the fields galaxies clearly present a complex morphologies (e.g. G1\_1 and G1\_3). For each isolated source, we determined properties using GALFIT, assuming Sersic profiles. In Tab. \ref{tab_galaxies}, we report the apparent model magnitude (together with the aperture one), the Sersic index, the effective radius R$_{eff}$ and the ellipticity.
For some of the field galaxies, the fit with GALFIT was not possible. This is probably due to the presence of very structured source (e.g. G1\_1) or source with close companion (e.g. G1\_3).
 
Despite the modest seeing conditions (Sect. \ref{sect_MADobs}), we are able to characterize the morphology of very faint and small field galaxies with Ks apparent magnitude in the range 16--18 and sub-arcsecond scale length, demonstrating the final good potentialities of these MAD data.  

\begin{table}
\caption{Properties of the faint galaxies in the QSOs fields.} \label{tab_galaxies}
\centering
\begin{tabular}{|lrrrlllll|}
\hline
QSO   & Galaxy &  $\Delta$RA 	&  $\Delta$DEC  &m(K)$_{ap}$ & m(K)$_{md}$  & R$_{eff}$      & n & e \\
field & ID            &  arcsec   &    arcsec                &                		&        		& arcsec         &   &    \\
\hline\hline
QSO\#1 	& G1\_1   &  14.79 &  - 3.30 & 17.5  & --    & --     & --      &  --    \\
 	         & G1\_2   & 12.47 & 2.55 & 18.0  & 17.7 & 0.3    & 1.8     & 0.5 \\
 		& G1\_3   & 4.29 & 7.58 & 16.2  & --    & --     & --     &   --   \\ 
QSO\#2	& G2\_1	& 10.55 & 16.69   & 17.4    & 16.7 & 1.0 &  7 & 0.4  \\
QSO\#5  & G5\_1& 7.62& -9.22 & 17.6    & 17.3 & 0.8 & 2.0  & 0.5  \\
	& G5\_2	&-0.63& -4.03& 17.2    & --   & -- & --   & --  \\
QSO\#6  & G6\_1 &15.23&4.57   &16.9   & 16.8 & 0.2  & 1.5 & 0.1   \\
	& G6\_2 &14.47  & 4.19&17.8   & 17.2 & 0.4  & 2.0 & 0.3  \\
	& G6\_3 &  1.35& 1.49& 18.0   & 17.6 & 0.3  & 3.2 & 0.1   \\
	& G6\_4 &-12.44  & -13.18& 17.2   & 16.4 & 1.0  & 4.2 & 0.4   \\
\hline
\end{tabular}
\tablenotetext{}{
\footnotesize Col.1: QSO field;  col.2: field galaxy's ID;  col. 3-4: angular distance in arcsec from the QSO; col. 5: aperture apparent magnitude in Ks band; col.6: apparent model magnitude; col. 7: effective radius in arcsec; col.8: Sersic index; col. 9: ellipticity.}
\end{table}

\end{document}